\documentclass[a4paper,11pt]{article}
\usepackage[utf8]{inputenc}
\usepackage[top=0.95in, bottom=0.75in, left=1.28in, right=0.8in]{geometry}
\usepackage{mathtools,hyperref}
\usepackage{tikz}
\usepackage{graphics}
\usepackage{setspace}
\title{\large \textbf {Interaction of Subharmonic Light Modes with Three-Level Atom}}
\author{\large Merid Tufa and Fesseha Kassahun\\Department of Physics, Addis Ababa University, P. O. Box 1176, Addis Ababa, Ethiopia}

\begin{document}
\maketitle
\begin{abstract}
We have considered the interaction of the subharmonic light modes with a three-level atom. We have found that the effect of this interaction is to decrease the quadrature squeezing and the mean photon number of the two-mode cavity light.
\end{abstract}
\providecommand{\keywords}[1]{\textbf{{keywords:}} #1}
\keywords{Quadrature squeezing, Mean photon number, Subharmonic light modes}
\section{Introduction}
Squeezing is one of the nonclassical features of light that has attracted a great deal of interest. Several authors have carried out the analysis of the quantum properties of the squeezed light generated by various optical systems [1-10]. In squeezed light the noise in one quadrature is below the vacuum-state level with the product of the uncertainties in the two quadratures satisfying the uncertainty relation [9-11]. Squeezed light has potential applications in low-noise optical communications, precision measurements, and weak signal detections [2, 3].\\
\indent
Fesseha [9,10] has studied the squeezing and statistical properties of the light produced by three-level atoms in a closed cavity and pumped by electron bombardment at a constant rate. He has found that the generated light is in a squeezed state, with the maximum quadrature squeezing being $50\%$ below the vacuum-state level. In addition he has considered the interaction of three-level atoms pumped by coherent light, with the maximum quadrature squeezing being $43.4\%$ below the vacuum-state level.\\ 
\indent
On the other hand, it has been shown theoretically [10-13] and subsequently confirmed experimentally [14-16] that in a subharmonic generator a nonlinear crystal pumped by coherent light produces squeezed subharmonic light modes. It is found that the superposition of this modes has a maximum squeezing of $50\%$ below the vacuum-state level [10]. In addition, some authors have studied the statistical and squeezing properties of the light generated by three-level atoms interacting with subharmonic light modes, using the usual commutation relation [17-20]. However, it appears to be difficult to believe the results obtained in this manner to be correct.\\
\indent
Here we  have considered a revised version of [21] which deals with the interaction of a three-level atom with the subharmonic light modes emerging from a nonlinear crystal pumped by coherent light and available in a closed cavity coupled to a vacuum reservoir. Our interest is to analyze the squeezing and statistical properties of the cavity modes available following this interaction. We carry out our analysis applying the master equation for the cavity modes and atomic operators. The large-time approximation is used to decouple the equations of evolution of cavity modes and atomic operators. Finally, employing the steady-state solutions of the resulting equations of evolution for the expectation values of the cavity modes and atomic operators, we calculate the mean photon number and the quadrature squeezing.\\

\section{Operator Dynamics}
\noindent 
We consider the case in which the top and bottom levels of the three-level atom are not coupled by the coherent light emerging from the nonlinear crystal. This is physically realized by covering the right-side of the nonlinear crystal by a screen which can absorb the coherent light. We denote the top, intermediate, and bottom levels of the atom by $ | a\rangle, |b\rangle$ and $ |c\rangle$, respectively.\\ 
\indent
The process of subharmonic generation taking place inside the nonlinear crystal can be described by the  Hamiltonian
\begin{equation}\label{1}
 \hat{H}' = i\lambda(\hat{c}^{\dagger}\hat{a}\hat{b} - \hat{c}\hat{a}^{\dagger}\hat{b}^{\dagger}),
\end{equation}
where the operators $\hat{a}$ and $\hat{b}$ represent the subharmonic light modes, $\lambda$ is the coupling constant between the coherent light and light mode a or b and $\hat{c}$ is the annihilation operator for the coherent light. Upon replacing the operator $\hat{c}$ by $\gamma$ which is taken to be real, positive, and constant, we can write the Hamiltonian as
\begin{equation}\label{2}
 \hat{H}' = i\varepsilon(\hat{a}\hat{b} - \hat{a}^{\dagger}\hat{b}^{\dagger}),
\end{equation}
where $ \varepsilon = \lambda\gamma$.
\begin{figure*}
 \centering
 \begin{center}
 \includegraphics[width=8cm, height = 5cm]{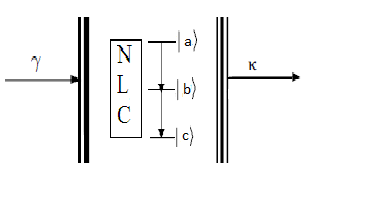}
  \caption{A three-level atom with a nonlinear crystal (NLC).}
 \end{center}
\end{figure*}
 In addition, the interaction of the subharmonic light modes with the three-level atom can be described at resonance by the Hamiltonian
\begin{equation}\label{3}
 \hat{H}'' = ig(\hat{\sigma}^{\dagger}_a\hat{a} - \hat{a}^{\dagger}\hat{\sigma}_a + \hat{\sigma}^{\dagger}_b\hat{b} - \hat{b}^{\dagger}\hat{\sigma}_b),
\end{equation}
where $g$ is the coupling constant between the atom and light mode a or b and $\hat{\sigma}_a$ and $\hat{\sigma}_b$ are lowering atomic operators defined by
\begin{equation}\label{4}
 \hat{\sigma}_a = |b\rangle\langle a|,
\end{equation}
\begin{equation}\label{5}
 \hat{\sigma}_b = |c\rangle\langle b|.
\end{equation}
\noindent
Thus the interactions involving the cavity light modes are described by
\begin{equation}\label{6}
 \hat{H_c} = i\varepsilon(\hat{a}\hat{b} - \hat{a}^{\dagger}\hat{b}^{\dagger}) + ig(\hat{\sigma}^{\dagger}_a\hat{a} - \hat{a}^{\dagger}\hat{\sigma}_a + \hat{\sigma}^{\dagger}_b\hat{b} - \hat{b}^{\dagger}\hat{\sigma}_b)
\end{equation}
and the interaction involving the three-atom are given by
\begin{equation}\label{7}
\hat{H}_a = ig(\hat{\sigma}^{\dagger}_a\hat{a} - \hat{a}^{\dagger}\hat{\sigma}_a + \hat{\sigma}^{\dagger}_b\hat{b} - \hat{b}^{\dagger}\hat{\sigma}_b).
\end{equation}
\indent
Now we seek to obtain the equations of evolution for the cavity mode operators employing the master equation 
\begin{equation}\label{8}
 \frac{d}{dt}\hat{\rho}(t) = - i\big[\hat{H}_c, \hat{\rho}\big]  + \frac{\kappa}{2}\bigg(2\hat{a}\rho\hat{a}^{\dagger} - \hat{a}^{\dagger}\hat{a}\rho - \rho\hat{a}^{\dagger}\hat{a}\bigg) + \frac{\kappa}{2}\bigg(2\hat{b}\rho\hat{b}^{\dagger} - \hat{b}^{\dagger}\hat{b}\rho - \rho\hat{b}^{\dagger}\hat{b}\bigg).
\end{equation}
On account of Eq. \eqref{6}, we have
\begin{eqnarray}\label{9}
 \frac{d}{dt}\hat{\rho}(t) &=& - i\big[i\varepsilon(\hat{a}\hat{b} - \hat{a}^{\dagger}\hat{b}^{\dagger}) + ig(\hat{\sigma}^{\dagger}_a\hat{a} - \hat{a}^{\dagger}\hat{\sigma}_a + \hat{\sigma}^{\dagger}_b\hat{b} - \hat{b}^{\dagger}\hat{\sigma}_b), \hat{\rho}\big]  + \frac{\kappa}{2}\bigg(2\hat{a}\rho\hat{a}^{\dagger} - \hat{a}^{\dagger}\hat{a}\rho\nonumber\\
 &-& \rho\hat{a}^{\dagger}\hat{a}\bigg) + \frac{\kappa}{2}\bigg(2\hat{b}\rho\hat{b}^{\dagger} - \hat{b}^{\dagger}\hat{b}\rho - \rho\hat{b}^{\dagger}\hat{b}\bigg).
\end{eqnarray}
We recall that the equation of evolution for an operator $\hat{A}$ is expressible as
\begin{equation}\label{10}
 \frac{d}{dt}\langle\hat{A}\rangle = Tr\bigg(\frac{d}{dt}\hat{\rho}(t)\hat{A}\bigg).
\end{equation}
Then employing this relation along with Eq. \eqref{9}, we readily obtain
\begin{equation}\label{11}
 \frac{d}{dt}\langle\hat{a}(t)\rangle = -\frac{\kappa}{2}\langle\hat{a}(t)\rangle - \varepsilon\langle\hat{b}^{\dagger}(t)\rangle - g\langle\hat{\sigma}_a(t)\rangle,
\end{equation}
\begin{equation}\label{12}
 \frac{d}{dt}\langle\hat{b}(t)\rangle = -\frac{\kappa}{2}\langle\hat{b}(t)\rangle - \varepsilon\langle\hat{a}^{\dagger}(t)\rangle - g\langle\hat{\sigma}_b(t)\rangle,
\end{equation}
\begin{equation}\label{13}
 \frac{d}{dt}\langle\hat{a}^{\dagger}(t)\hat{a}(t)\rangle = -\kappa\langle\hat{a}^{\dagger}(t)\hat{a}(t)\rangle - \varepsilon\big\langle\hat{b}(t)\hat{a}(t) +\hat{a}^{\dagger}(t)\hat{b}^{\dagger}(t)\big\rangle - g\big\langle\hat{\sigma}^{\dagger}_a(t)\hat{a}(t) + \hat{a}^{\dagger}(t)\hat{\sigma}_a(t)\big\rangle,
\end{equation}
\begin{equation}\label{14}
 \frac{d}{dt}\langle\hat{a}(t)\hat{a}^{\dagger}(t)\rangle = -\kappa\langle\hat{a}(t)\hat{a}^{\dagger}(t)\rangle - \varepsilon\big\langle\hat{a}(t)\hat{b}(t) +\hat{b}^{\dagger}(t)\hat{a}^{\dagger}(t)\big\rangle - g\big\langle\hat{a}(t)\hat{\sigma}^{\dagger}_a(t) + \hat{\sigma}_a(t)\hat{a}^{\dagger}(t)\big\rangle + \kappa,
\end{equation}
\begin{equation}\label{15}
 \frac{d}{dt}\langle\hat{b}^{\dagger}(t)\hat{b}(t)\rangle = -\kappa\langle\hat{b}^{\dagger}(t)\hat{b}(t)\rangle - \varepsilon\big\langle\hat{a}(t)\hat{b}(t) +\hat{b}^{\dagger}(t)\hat{a}^{\dagger}(t)\big\rangle - g\big\langle\hat{\sigma}^{\dagger}_b(t)\hat{b}(t) + \hat{b}^{\dagger}(t)\hat{\sigma}_b(t)\big\rangle,
\end{equation}
\begin{equation}\label{16}
 \frac{d}{dt}\langle\hat{b}(t)\hat{b}^{\dagger}(t)\rangle = -\kappa\langle\hat{b}(t)\hat{b}^{\dagger}(t)\rangle - \varepsilon\big\langle\hat{b}(t)\hat{a}(t) +\hat{a}^{\dagger}(t)\hat{b}^{\dagger}(t)\big\rangle - g\big\langle\hat{b}(t)\hat{\sigma}^{\dagger}_b(t) + \hat{\sigma}_b(t)\hat{b}^{\dagger}(t)\big\rangle + \kappa,
\end{equation}
\begin{equation}\label{17}
\frac{d}{dt}\langle\hat{a}(t)\hat{b}(t)\rangle = -\kappa\langle\hat{a}(t)\hat{b}(t)\rangle - \varepsilon\big\langle\hat{a}^{\dagger}(t)\hat{a}(t) +\hat{b}^{\dagger}(t)\hat{b}(t)\big\rangle - g\big\langle\hat{\sigma}_a(t)\hat{b}(t) + \hat{a}(t)\hat{\sigma}_b(t)\big\rangle - \varepsilon,
\end{equation}
\begin{equation}\label{18}
\frac{d}{dt}\langle\hat{b}(t)\hat{a}(t)\rangle = -\kappa\langle\hat{b}(t)\hat{a}(t)\rangle - \varepsilon\big\langle\hat{a}^{\dagger}(t)\hat{a}(t) +\hat{b}^{\dagger}(t)\hat{b}(t)\big\rangle - g\big\langle\hat{b}(t)\hat{\sigma}_a(t) + \hat{\sigma}_b(t)\hat{a}(t)\big\rangle - \varepsilon,
\end{equation}
\begin{equation}\label{19}
\frac{d}{dt}\langle\hat{a}^{2}(t)\rangle = -\kappa\langle\hat{a}^2(t)\rangle - \varepsilon\big\langle\hat{a}(t)\hat{b}^{\dagger}(t) +\hat{b}^{\dagger}(t)\hat{a}(t)\big\rangle - g\big\langle\hat{a}(t)\hat{\sigma}_a(t) + \hat{\sigma}_a(t)\hat{a}(t)\big\rangle,
\end{equation}
\begin{equation}\label{20}
\frac{d}{dt}\langle\hat{b}^{2}(t)\rangle = -\kappa\langle\hat{b}^2(t)\rangle - \varepsilon\big\langle\hat{b}(t)\hat{a}^{\dagger}(t) +\hat{a}^{\dagger}(t)\hat{b}(t)\big\rangle - g\big\langle\hat{b}(t)\hat{\sigma}_b(t) + \hat{\sigma}_b(t)\hat{b}(t)\big\rangle,
\end{equation}
\begin{equation}\label{21}
 \frac{d}{dt}\langle\hat{a}^{\dagger}(t)\hat{b}(t)\rangle = -\kappa\langle\hat{a}^{\dagger}(t)\hat{b}(t)\rangle - \varepsilon\big\langle\hat{a}^{\dagger 2}(t) +\hat{b}^{2}(t)\big\rangle - g\big\langle\hat{\sigma}^{\dagger}_a(t)\hat{b}(t) + \hat{a}^{\dagger}(t)\hat{\sigma}_b(t)\big\rangle,
\end{equation}
and
\begin{equation}\label{22}
 \frac{d}{dt}\langle\hat{b}(t)\hat{a}^{\dagger}(t)\rangle = -\kappa\langle\hat{b}(t)\hat{a}^{\dagger}(t)\rangle - \varepsilon\big\langle\hat{a}^{\dagger 2}(t) +\hat{b}^{2}(t)\big\rangle - g\big\langle\hat{b}(t)\hat{\sigma}^{\dagger}_a(t) + \hat{\sigma}_b(t)\hat{a}^{\dagger}(t)\big\rangle.
\end{equation}
One can rewrite Eqs. \eqref{11} and \eqref{12} as 
\begin{equation}\label{23}
 \frac{d}{dt}\hat{a}(t) = -\frac{\kappa}{2}\hat{a}(t) -\varepsilon\hat{b}^{\dagger}(t)
 - g\hat{\sigma}_a(t) + \hat{F}_a(t),
\end{equation}
\begin{equation}\label{24}
\frac{d}{dt}\hat{b}(t) = -\frac{\kappa}{2}\hat{b}(t) -\varepsilon\hat{a}^{\dagger}(t)
 - g\hat{\sigma}_b(t) + \hat{F}_b(t),
\end{equation}
where $\hat{F}_a(t)$ and $\hat{F}_b(t)$ are noise operators with vanishing mean and associated with the cavity mode operators $\hat{a}(t)$ and $\hat{b}(t)$, respectively.\\
\indent
We see that Eqs. \eqref{13}-\eqref{22} are nonlinear differential equations and hence it is not possible to find the exact time-dependent solutions of these equations. To overcome this problem, we apply the large-time approximation scheme to Eqs. \eqref{23} and \eqref{24} and get the following approximately valid relations
\begin{equation}\label{25}
 \hat{a}(t) = -\frac{2\varepsilon}{\kappa}\hat{b}^{\dagger}(t) - \frac{2g}{\kappa}\hat{\sigma}_a(t) + \frac{2}{\kappa}\hat{F}_a(t)
\end{equation}
and
\begin{equation}\label{26}
 \hat{b}(t) = -\frac{2\varepsilon}{\kappa}\hat{a}^{\dagger}(t) - \frac{2g}{\kappa}\hat{\sigma}_b(t) + \frac{2}{\kappa}\hat{F}_b(t).
\end{equation}
It then follows that
\begin{equation}\label{27}
 \hat{a}(t) = \frac{4\varepsilon\kappa g}{\kappa^2 - 4\varepsilon^2}\bigg(\frac{\hat{\sigma}^{\dagger}_b(t)}{\kappa} - \frac{\hat{\sigma}_a(t)}{2\varepsilon} + \frac{\hat{F}_a(t)}{2\varepsilon g} - \frac{\hat{F}^{\dagger}_b(t)}{\kappa g}\bigg)
\end{equation}
and
\begin{equation}\label{28}
 \hat{b}(t) = \frac{4\varepsilon\kappa g}{\kappa^2 - 4\varepsilon^2}\bigg(\frac{\hat{\sigma}^{\dagger}_a(t)}{\kappa} - \frac{\hat{\sigma}_b(t)}{2\varepsilon} + \frac{\hat{F}_b(t)}{2\varepsilon g} - \frac{\hat{F}^{\dagger}_a(t)}{\kappa g}\bigg).
\end{equation}
Upon substituting Eqs. \eqref{27} and \eqref{28} together with their adjoint into Eqs. \eqref{13}-\eqref{22}, we readily obtain
\begin{eqnarray}\label{29}
 \frac{d}{dt}\langle\hat{a}^{\dagger}(t)\hat{a}(t)\rangle &=& -\kappa\langle\hat{a}^{\dagger}(t)\hat{a}(t)\rangle - \varepsilon\big\langle\hat{b}(t)\hat{a}(t) +\hat{a}^{\dagger}(t)\hat{b}^{\dagger}(t)\big\rangle - \frac{4\varepsilon\kappa g^2}{\kappa^2 - 4\varepsilon^2}\bigg[\frac{\langle\hat{\sigma}_c(t) + \hat{\sigma}^{\dagger}_c(t)\rangle}{\kappa}\nonumber\\ 
 &&- \frac{\langle\hat{\eta}_a(t)\rangle}{\varepsilon} + \bigg\langle\hat{\sigma}^{\dagger}_a(t) \bigg(\frac{\hat{F}_a(t)}{2\varepsilon g} - \frac{\hat{F}^{\dagger}_b(t)}{\kappa g}\bigg) + \bigg(\frac{\hat{F}^{\dagger}_a(t)}{2\varepsilon g} - \frac{\hat{F}_b(t)}{\kappa g}\bigg)\hat{\sigma}_a(t)\bigg\rangle\bigg],
\end{eqnarray}
\begin{eqnarray}\label{30}
 \frac{d}{dt}\langle\hat{a}(t)\hat{a}^{\dagger}(t)\rangle &=& -\kappa\langle\hat{a}(t)\hat{a}^{\dagger}(t)\rangle - \varepsilon\big\langle\hat{a}(t)\hat{b}(t) +\hat{b}^{\dagger}(t)\hat{a}^{\dagger}(t)\big\rangle - \frac{4\varepsilon\kappa g^2}{\kappa^2 - 4\varepsilon^2}\bigg[-\frac{\langle\hat{\eta}_b(t)\rangle}{\varepsilon}\nonumber\\
 &&+ \bigg\langle\bigg(\frac{\hat{F}_a(t)}{2\varepsilon g} - \frac{\hat{F}^{\dagger}_b(t)}{\kappa g}\bigg)\hat{\sigma}^{\dagger}_a(t)  + \hat{\sigma}_a(t)\bigg(\frac{\hat{F}^{\dagger}_a(t)}{2\varepsilon g} - \frac{\hat{F}_b(t)}{\kappa g}\bigg)\bigg\rangle\bigg] + \kappa,
\end{eqnarray}
\begin{eqnarray}\label{31}
 \frac{d}{dt}\langle\hat{b}^{\dagger}(t)\hat{b}(t)\rangle &=& -\kappa\langle\hat{b}^{\dagger}(t)\hat{b}(t)\rangle - \varepsilon\big\langle\hat{a}(t)\hat{b}(t) +\hat{b}^{\dagger}(t)\hat{a}^{\dagger}(t)\big\rangle - \frac{4\varepsilon\kappa g^2}{\kappa^2 - 4\varepsilon^2}\bigg[-\frac{\langle\hat{\eta}_b(t)\rangle}{\varepsilon}\nonumber\\
&&+ \bigg\langle\hat{\sigma}^{\dagger}_b(t) \bigg(\frac{\hat{F}_b(t)}{2\varepsilon g} - \frac{\hat{F}^{\dagger}_a(t)}{\kappa g}\bigg) + \bigg(\frac{\hat{F}^{\dagger}_b(t)}{2\varepsilon g} - \frac{\hat{F}_a(t)}{\kappa g}\bigg)\hat{\sigma}_b(t)\bigg\rangle\bigg],
\end{eqnarray}
\begin{eqnarray}\label{32}
 \frac{d}{dt}\langle\hat{b}(t)\hat{b}^{\dagger}(t)\rangle &=& -\kappa\langle\hat{b}(t)\hat{b}^{\dagger}(t)\rangle - \varepsilon\big\langle\hat{b}(t)\hat{a}(t) +\hat{a}^{\dagger}(t)\hat{b}^{\dagger}(t)\big\rangle - \frac{4\varepsilon\kappa g^2}{\kappa^2 - 4\varepsilon^2}\bigg[\frac{\langle\hat{\sigma}_c(t) + \hat{\sigma}^{\dagger}_c(t)\rangle}{\kappa}\nonumber\\
 &&- \frac{\langle\hat{\eta}_c(t)\rangle}{\varepsilon} + \bigg\langle\bigg(\frac{\hat{F}_b(t)}{2\varepsilon g} - \frac{\hat{F}^{\dagger}_a(t)}{\kappa g}\bigg)\hat{\sigma}^{\dagger}_b(t)  + \hat{\sigma}_b(t)\bigg(\frac{\hat{F}^{\dagger}_b(t)}{2\varepsilon g} - \frac{\hat{F}_a(t)}{\kappa g}\bigg)\bigg\rangle\bigg]\nonumber\\&& + \kappa,
\end{eqnarray}
\begin{eqnarray}\label{33}
\frac{d}{dt}\langle\hat{a}(t)\hat{b}(t)\rangle &=& -\kappa\langle\hat{a}(t)\hat{b}(t)\rangle - \varepsilon\big\langle\hat{a}^{\dagger}(t)\hat{a}(t) +\hat{b}^{\dagger}(t)\hat{b}(t)\big\rangle - \frac{4\varepsilon\kappa g^2}{\kappa^2 - 4\varepsilon^2}\bigg[\frac{2\langle\hat{\eta}_b(t)\rangle}{\kappa}\nonumber\\
&&+ \bigg\langle\hat{\sigma}_a(t) \bigg(\frac{\hat{F}_b(t)}{2\varepsilon g} - \frac{\hat{F}^{\dagger}_a(t)}{\kappa g}\bigg) + \bigg(\frac{\hat{F}_a(t)}{2\varepsilon g} - \frac{\hat{F}^{\dagger}_b(t)}{\kappa g}\bigg)\hat{\sigma}_b(t)\bigg\rangle\bigg] -\varepsilon,
\end{eqnarray}
\begin{eqnarray}\label{34}
\frac{d}{dt}\langle\hat{b}(t)\hat{a}(t)\rangle &=& -\kappa\langle\hat{b}(t)\hat{a}(t)\rangle - \varepsilon\big\langle\hat{a}^{\dagger}(t)\hat{a}(t) +\hat{b}^{\dagger}(t)\hat{b}(t)\big\rangle - \frac{4\varepsilon\kappa g^2}{\kappa^2 - 4\varepsilon^2}\bigg[\frac{\langle\hat{\eta}_a(t) + \hat{\eta}_c(t)\rangle}{\kappa}\nonumber\\
 &&- \frac{\langle\hat{\sigma}_c(t)\rangle}{\varepsilon} + \bigg\langle\bigg(\frac{\hat{F}_b(t)}{2\varepsilon g} - \frac{\hat{F}^{\dagger}_a(t)}{\kappa g}\bigg)\hat{\sigma}_a(t)  + \hat{\sigma}_b(t)\bigg(\frac{\hat{F}_a(t)}{2\varepsilon g} - \frac{\hat{F}^{\dagger}_b(t)}{\kappa g}\bigg)\bigg\rangle\bigg]\nonumber\\&&  -\varepsilon,
\end{eqnarray}
\begin{eqnarray}\label{35}
\frac{d}{dt}\langle\hat{a}^{2}(t)\rangle&=&-\kappa\langle\hat{a}^2(t)\rangle - \varepsilon\big\langle\hat{a}(t)\hat{b}^{\dagger}(t) +\hat{b}^{\dagger}(t)\hat{a}(t)\big\rangle - \frac{4\varepsilon\kappa g^2}{\kappa^2 - 4\varepsilon^2}\bigg[\bigg\langle\bigg(\frac{\hat{F}_a(t)}{2\varepsilon g} - \frac{\hat{F}^{\dagger}_b(t)}{\kappa g}\bigg)\hat{\sigma}_a(t)\nonumber\\&&+\hat{\sigma}_a(t)\bigg(\frac{\hat{F}_a(t)}{2\varepsilon g} - \frac{\hat{F}^{\dagger}_b(t)}{\kappa g}\bigg)\bigg\rangle\bigg],
\end{eqnarray}
\begin{eqnarray}\label{36}
\frac{d}{dt}\langle\hat{b}^{2}(t)\rangle &=& -\kappa\langle\hat{b}^2(t)\rangle - \varepsilon\big\langle\hat{b}(t)\hat{a}^{\dagger}(t) +\hat{a}^{\dagger}(t)\hat{b}(t)\big\rangle - \frac{4\varepsilon\kappa g^2}{\kappa^2 - 4\varepsilon^2}\bigg[\bigg\langle \bigg(\frac{\hat{F}_b(t)}{2\varepsilon g} - \frac{\hat{F}^{\dagger}_a(t)}{\kappa g}\bigg)\hat{\sigma}_b(t)\nonumber\\
&&+\hat{\sigma}_b(t)\bigg(\frac{\hat{F}_b(t)}{2\varepsilon g} - \frac{\hat{F}^{\dagger}_a(t)}{\kappa g}\bigg)\bigg\rangle\bigg],
\end{eqnarray}
\begin{eqnarray}\label{37}
 \frac{d}{dt}\langle\hat{a}^{\dagger}(t)\hat{b}(t)\rangle &=& -\kappa\langle\hat{a}^{\dagger}(t)\hat{b}(t)\rangle - \varepsilon\big\langle\hat{a}^{\dagger 2}(t) +\hat{b}^{2}(t)\big\rangle - \frac{4\varepsilon\kappa g^2}{\kappa^2 - 4\varepsilon^2}\bigg[\bigg\langle \hat{\sigma}^{\dagger}_a(t) \bigg(\frac{\hat{F}_b(t)}{2\varepsilon g} - \frac{\hat{F}^{\dagger}_a(t)}{\kappa g}\bigg)\nonumber\\
 &&+ \bigg(\frac{\hat{F}^{\dagger}_a(t)}{2\varepsilon g} - \frac{\hat{F}_b(t)}{\kappa g}\bigg)\hat{\sigma}_b(t)\bigg\rangle\bigg],
\end{eqnarray}
\begin{eqnarray}\label{38}
 \frac{d}{dt}\langle\hat{b}(t)\hat{a}^{\dagger}(t)\rangle &=& -\kappa\langle\hat{b}(t)\hat{a}^{\dagger}(t)\rangle - \varepsilon\big\langle\hat{a}^{\dagger 2}(t) +\hat{b}^{2}(t)\big\rangle - \frac{4\varepsilon\kappa g^2}{\kappa^2 - 4\varepsilon^2}\bigg[\bigg\langle \bigg(\frac{\hat{F}_b(t)}{2\varepsilon g} - \frac{\hat{F}^{\dagger}_a(t)}{\kappa g}\bigg)\hat{\sigma}^{\dagger}_a(t) \nonumber\\
 &&+\hat{\sigma}_b(t) \bigg(\frac{\hat{F}^{\dagger}_a(t)}{2\varepsilon g} - \frac{\hat{F}_b(t)}{\kappa g}\bigg)\bigg\rangle\bigg],
\end{eqnarray}
where
\begin{equation}\label{39}
 \hat{\sigma}_c = |c\rangle\langle a|,
\end{equation}
\begin{equation}\label{40}
 \hat{\eta}_a = |a\rangle\langle a|,
\end{equation}
\begin{equation}\label{41}
 \hat{\eta}_b = |b\rangle\langle b|,
\end{equation}
and
\begin{equation}\label{42}
 \hat{\eta}_c = |c\rangle\langle c|.
\end{equation}
Since an atomic operator and a noise operator associated with a cavity mode are not correlated, one can write
\begin{equation}\label{43}
 \langle\hat{F}(t)\hat{\sigma}(t)\rangle = \langle\hat{F}(t)\rangle\langle\hat{\sigma}(t)\rangle = 0
\end{equation}
and hence Eqs. \eqref{29}-\eqref{38} take the form
\begin{eqnarray}\label{44}
 \frac{d}{dt}\langle\hat{a}^{\dagger}(t)\hat{a}(t)\rangle &=& -\kappa\langle\hat{a}^{\dagger}(t)\hat{a}(t)\rangle - \varepsilon\big\langle\hat{b}(t)\hat{a}(t) +\hat{a}^{\dagger}(t)\hat{b}^{\dagger}(t)\big\rangle - \frac{4\varepsilon\kappa g^2}{\kappa^2 - 4\varepsilon^2}\bigg[\frac{\langle\hat{\sigma}_c(t) + \hat{\sigma}^{\dagger}_c(t)\rangle}{\kappa} \nonumber\\&&- \frac{\langle\hat{\eta}_a(t)\rangle}{\varepsilon}\bigg],
\end{eqnarray}
\begin{eqnarray}\label{45}
 \frac{d}{dt}\langle\hat{a}(t)\hat{a}^{\dagger}(t)\rangle &=& -\kappa\langle\hat{a}(t)\hat{a}^{\dagger}(t)\rangle - \varepsilon\big\langle\hat{a}(t)\hat{b}(t) +\hat{b}^{\dagger}(t)\hat{a}^{\dagger}(t)\big\rangle + \frac{4\varepsilon\kappa g^2}{\kappa^2 - 4\varepsilon^2}\frac{\langle\hat{\eta}_b(t)\rangle}{\varepsilon}\nonumber\\
 &&+ \kappa,
\end{eqnarray}
\begin{eqnarray}\label{46}
 \frac{d}{dt}\langle\hat{b}^{\dagger}(t)\hat{b}(t)\rangle = -\kappa\langle\hat{b}^{\dagger}(t)\hat{b}(t)\rangle - \varepsilon\big\langle\hat{a}(t)\hat{b}(t) +\hat{b}^{\dagger}(t)\hat{a}^{\dagger}(t)\big\rangle + \frac{4\varepsilon\kappa g^2}{\kappa^2 - 4\varepsilon^2}\frac{\hat{\langle\eta}_b(t)\rangle}{\varepsilon},
\end{eqnarray}
\begin{eqnarray}\label{47}
 \frac{d}{dt}\langle\hat{b}(t)\hat{b}^{\dagger}(t)\rangle &=& -\kappa\langle\hat{b}(t)\hat{b}^{\dagger}(t)\rangle - \varepsilon\big\langle\hat{b}(t)\hat{a}(t) +\hat{a}^{\dagger}(t)\hat{b}^{\dagger}(t)\big\rangle 
 - \frac{4\varepsilon\kappa g^2}{\kappa^2 - 4\varepsilon^2}\bigg[\frac{\langle\hat{\sigma}_c(t) + \hat{\sigma}^{\dagger}_c(t)\rangle}{\kappa}\nonumber\\&& - \frac{\langle\hat{\eta}_c(t)\rangle}{\varepsilon}\bigg] + \kappa,
\end{eqnarray}
\begin{eqnarray}\label{48}
\frac{d}{dt}\langle\hat{a}(t)\hat{b}(t)\rangle &=& -\kappa\langle\hat{a}(t)\hat{b}(t)\rangle - \varepsilon\big\langle\hat{a}^{\dagger}(t)\hat{a}(t) +\hat{b}^{\dagger}(t)\hat{b}(t)\big\rangle - \frac{4\varepsilon\kappa g^2}{\kappa^2 - 4\varepsilon^2}\frac{2\langle\hat{\eta}_b(t)\rangle}{\kappa} \nonumber\\&&- \varepsilon,
\end{eqnarray}
\begin{eqnarray}\label{49}
\frac{d}{dt}\langle\hat{b}(t)\hat{a}(t)\rangle &=& -\kappa\langle\hat{b}(t)\hat{a}(t)\rangle - \varepsilon\big\langle\hat{a}^{\dagger}(t)\hat{a}(t) +\hat{b}^{\dagger}(t)\hat{b}(t)\big\rangle - \frac{4\varepsilon\kappa g^2}{\kappa^2 - 4\varepsilon^2}\bigg[\frac{\langle\hat{\eta}_a(t) + \hat{\eta}_c(t)\rangle}{\kappa}
 \nonumber\\&&- \frac{\langle\hat{\sigma}_c(t)\rangle}{\varepsilon}\bigg] -\varepsilon,
\end{eqnarray}
\begin{eqnarray}\label{50}
\frac{d}{dt}\langle\hat{a}^{2}(t)\rangle = -\kappa\langle\hat{a}^2(t)\rangle - \varepsilon\big\langle\hat{a}(t)\hat{b}^{\dagger}(t) +\hat{b}^{\dagger}(t)\hat{a}(t)\big\rangle,
\end{eqnarray}
\begin{eqnarray}\label{51}
\frac{d}{dt}\langle\hat{b}^{2}(t)\rangle = -\kappa\langle\hat{b}^2(t)\rangle - \varepsilon\big\langle\hat{b}(t)\hat{a}^{\dagger}(t) +\hat{a}^{\dagger}(t)\hat{b}(t)\big\rangle,
\end{eqnarray}
\begin{eqnarray}\label{52}
 \frac{d}{dt}\langle\hat{a}^{\dagger}(t)\hat{b}(t)\rangle = -\kappa\langle\hat{a}^{\dagger}(t)\hat{b}(t)\rangle - \varepsilon\big\langle\hat{a}^{\dagger 2}(t) +\hat{b}^{2}(t)\big\rangle.
\end{eqnarray}
and
\begin{eqnarray}\label{53}
 \frac{d}{dt}\langle\hat{b}(t)\hat{a}^{\dagger}(t)\rangle = -\kappa\langle\hat{b}(t)\hat{a}^{\dagger}(t)\rangle - \varepsilon\big\langle\hat{a}^{\dagger 2}(t) +\hat{b}^{2}(t)\big\rangle.
\end{eqnarray}
The steady-state solutions of Eqs. \eqref{44}-\eqref{53} are found to be
\begin{eqnarray}\label{54}
\langle\hat{a}^{\dagger}\hat{a}\rangle = -\frac{\varepsilon}{\kappa}\big\langle\hat{b}\hat{a} + \hat{a}^{\dagger}\hat{b}^{\dagger}\big\rangle - \frac{4g^2}{\kappa^2 - 4\varepsilon^2}\bigg[\frac{\varepsilon\langle\hat{\sigma}_c + \hat{\sigma}^{\dagger}_c\rangle}{\kappa} - \langle\hat{\eta}_a\rangle\bigg],
\end{eqnarray}
\begin{eqnarray}\label{55}
\langle\hat{a}\hat{a}^{\dagger}\rangle = -\frac{\varepsilon}{\kappa}\big\langle\hat{a}\hat{b} +\hat{b}^{\dagger}\hat{a}^{\dagger}\big\rangle + \frac{4g^2}{\kappa^2 - 4\varepsilon^2}\langle\hat{\eta}_b\rangle +1,
\end{eqnarray}
\begin{eqnarray}\label{56}
 \langle\hat{b}^{\dagger}\hat{b}\rangle = -\frac{\varepsilon}{\kappa}\big\langle\hat{a}\hat{b} +\hat{b}^{\dagger}\hat{a}^{\dagger}\big\rangle + \frac{4g^2}{\kappa^2 - 4\varepsilon^2}\langle\hat{\eta}_b\rangle,
\end{eqnarray}
\begin{eqnarray}\label{57}
\langle\hat{b}\hat{b}^{\dagger}\rangle = -\frac{\varepsilon}{\kappa}\big\langle\hat{b}\hat{a} +\hat{a}^{\dagger}\hat{b}^{\dagger}\big\rangle - \frac{4g^2}{\kappa^2 - 4\varepsilon^2}\bigg[\frac{\varepsilon\langle\hat{\sigma}_c + \hat{\sigma}^{\dagger}_c\rangle}{\kappa} - \langle\hat{\eta}_c\rangle\bigg] + 1,
\end{eqnarray}
\begin{eqnarray}\label{58}
\langle\hat{a}\hat{b}\rangle = -\frac{\varepsilon}{\kappa}\big\langle\hat{a}^{\dagger}\hat{a} + \hat{b}^{\dagger}\hat{b}\big\rangle - \frac{4g^2}{\kappa^2 - 4\varepsilon^2}\frac{2\varepsilon\langle\hat{\eta}_b\rangle}{\kappa} - \frac{\varepsilon}{\kappa},
\end{eqnarray}
\begin{eqnarray}\label{59}
\langle\hat{b}\hat{a}\rangle = -\frac{\varepsilon}{\kappa}\big\langle\hat{a}^{\dagger}\hat{a} +\hat{b}^{\dagger}\hat{b}\big\rangle - \frac{4g^2}{\kappa^2 - 4\varepsilon^2}\bigg[\frac{\varepsilon\langle\hat{\eta}_a + \hat{\eta}_c\rangle}{\kappa}  -\langle\hat{\sigma}_c\rangle\bigg] -\frac{\varepsilon}{\kappa},
\end{eqnarray}
\begin{eqnarray}\label{60}
\langle\hat{a}^2\rangle = -\frac{\varepsilon}{\kappa}\big\langle\hat{a}\hat{b}^{\dagger} +\hat{b}^{\dagger}\hat{a}\big\rangle,
\end{eqnarray}
\begin{eqnarray}\label{61}
\langle\hat{b}^2\rangle = -\frac{\varepsilon}{\kappa}\big\langle\hat{b}\hat{a}^{\dagger} +\hat{a}^{\dagger}\hat{b}\big\rangle,
\end{eqnarray}
\begin{eqnarray}\label{62}
\langle\hat{a}^{\dagger}\hat{b}\rangle = -\frac{\varepsilon}{\kappa}\big\langle\hat{a}^{\dagger 2} +\hat{b}^{2}\big\rangle,
\end{eqnarray}
and 
\begin{eqnarray}\label{63}
\langle\hat{b}\hat{a}^{\dagger}\rangle = -\frac{\varepsilon}{\kappa}\big\langle\hat{a}^{\dagger 2} +\hat{b}^{2}\big\rangle.
\end{eqnarray}
Taking the complex conjugate of Eq. \eqref{60}, we have
\begin{eqnarray}\label{64}
\langle\hat{a}^{\dagger2}\rangle = -\frac{\varepsilon}{\kappa}\big\langle\hat{b}\hat{a}^{\dagger} +\hat{a}^{\dagger}\hat{b}\big\rangle.
\end{eqnarray}
In view of Eqs. \eqref{61} and \eqref{64}, Eq. \eqref{62} takes the form
\begin{eqnarray}\label{65}
\langle\hat{a}^{\dagger}\hat{b}\rangle = \frac{2\varepsilon^2}{\kappa^2}\big\langle\hat{b}\hat{a}^{\dagger} +\hat{a}^{\dagger}\hat{b}\big\rangle
\end{eqnarray} 
and also on account of Eqs. \eqref{62} and \eqref{63}, we have
\begin{eqnarray}\label{66}
\langle\hat{a}^{\dagger}\hat{b}\rangle = \frac{4\varepsilon^2}{\kappa^2}\big\langle\hat{a}^{\dagger}\hat{b}\big\rangle.
\end{eqnarray}  
This shows that
\begin{eqnarray}\label{67}
\langle\hat{a}^{\dagger}\hat{b}\rangle = \big\langle\hat{b}\hat{a}^{\dagger}\rangle = 0.
\end{eqnarray} 
In view of these results and their complex conjugates, Eqs. \eqref{60}, \eqref{61} and \eqref{64} turn out to be
\begin{eqnarray}\label{68}
\langle\hat{a}^2\rangle = \langle\hat{b}^2\rangle = \langle\hat{a}^{\dagger2}\rangle = \langle\hat{b}^{\dagger2}\rangle = 0.
\end{eqnarray}
Using Eqs.\eqref{58} and \eqref{59} along with their complex conjugates in Eqs. \eqref{54} and \eqref{55}, we easily find
\begin{eqnarray}\label{69}
 \langle\hat{a}^{\dagger}\hat{a}\rangle &=& \frac{2\varepsilon^2}{\kappa^2 - 2\varepsilon^2} + \frac{2\varepsilon^2}{\kappa^2 - 2\varepsilon^2}\langle\hat{b}^{\dagger}\hat{b}\rangle + \frac{4g^2}{(\kappa^2 - 2\varepsilon^2)(\kappa^2 - 4\varepsilon^2)} \bigg(2\varepsilon^2\langle\hat{\eta}_c\rangle + \langle\hat{\eta}_a\rangle(2\varepsilon^2 + \kappa^2)\nonumber\\&& - 2\varepsilon\kappa\big\langle\hat{\sigma}_c + \hat{\sigma}_c^{\dagger}\big\rangle\bigg)
\end{eqnarray}
and 
\begin{eqnarray}\label{70}
 \langle\hat{b}^{\dagger}\hat{b}\rangle &=& \frac{2\varepsilon^2}{\kappa^2 - 2\varepsilon^2} + \frac{2\varepsilon^2}{\kappa^2 - 2\varepsilon^2}\langle\hat{a}^{\dagger}\hat{a}\rangle + \frac{4g^2}{(\kappa^2 - 2\varepsilon^2)(\kappa^2 - 4\varepsilon^2)}\bigg(4\varepsilon^2 + \kappa^2\bigg)\langle\hat{\eta}_b\rangle.
\end{eqnarray}
Using Eqs. \eqref{69} and \eqref{70}, we readily obtain
\begin{eqnarray}\label{71}
 \langle\hat{a}^{\dagger}\hat{a}\rangle &=& \frac{2\varepsilon^2}{\kappa^2 - 4\varepsilon^2} + \frac{4g^2}{\kappa^2(\kappa^2 - 4\varepsilon^2)^2}\bigg((\kappa^4 - 4\varepsilon^4 )\langle\hat{\eta}_a\rangle + 2\varepsilon^2(4\varepsilon^2 + \kappa^2)\langle\hat{\eta}_b\rangle\nonumber\\&& +2\varepsilon^2(\kappa^2 - 2\varepsilon^2)\langle\hat{\eta}_c\rangle - 2\varepsilon\kappa(\kappa^2 - 2\varepsilon^2)\big\langle\hat{\sigma}_c + \hat{\sigma}^{\dagger}_c\big\rangle\bigg)
\end{eqnarray}
and 
\begin{eqnarray}\label{72}
 \langle\hat{b}^{\dagger}\hat{b}\rangle &=&\frac{2\varepsilon^2}{\kappa^2 - 4\varepsilon^2} + \frac{4g^2}{\kappa^2(\kappa^2 - 4\varepsilon^2)^2}\bigg(2\varepsilon^2(\kappa^2 + 2\varepsilon^2)\langle\hat{\eta}_a\rangle + (\kappa^2 + 4\varepsilon^2)(\kappa^2 - 2\varepsilon^2)\langle\hat{\eta}_b\rangle\nonumber\\&&+ 4\varepsilon^4\langle\hat{\eta}_c\rangle - 4\varepsilon^3\kappa\big\langle\hat{\sigma}_c + \hat{\sigma}^{\dagger}_c\big\rangle\bigg).
\end{eqnarray}
Furthermore, taking into account Eqs. \eqref{58} and \eqref{59} along with their complex conjugates, Eqs. \eqref{55} and \eqref{57} can be put in the form
\begin{eqnarray}\label{73}
\langle\hat{a}\hat{a}^{\dagger}\rangle &=& 1+ \frac{2\varepsilon^2}{\kappa^2 - 4\varepsilon^2} + \frac{4g^2}{\kappa^2(\kappa^2 - 4\varepsilon^2)^2}\bigg(2\varepsilon^2(\kappa^2 + 2\varepsilon^2)\langle\hat{\eta}_a\rangle + (\kappa^2 + 4\varepsilon^2)(\kappa^2 - 2\varepsilon^2)\langle\hat{\eta}_b\rangle\nonumber\\&&+ 4\varepsilon^4\langle\hat{\eta}_c\rangle - 4\varepsilon^3\kappa\big\langle\hat{\sigma}_c + \hat{\sigma}^{\dagger}_c\big\rangle\bigg)
\end{eqnarray}
and
\begin{eqnarray}\label{74}
\langle\hat{b}\hat{b}^{\dagger}\rangle &=& 1 +  \frac{2\varepsilon^2}{\kappa^2 - 4\varepsilon^2} + \frac{4g^2}{\kappa^2(\kappa^2 - 4\varepsilon^2)^2}\bigg(4\varepsilon^2(\kappa^2 - \varepsilon^2)\langle\hat{\eta}_a\rangle + 2\varepsilon^2(4\varepsilon^2 + \kappa^2)\langle\hat{\eta}_b\rangle\nonumber\\&& (\kappa^4 - 2\varepsilon^2(\kappa^2 + 2\varepsilon^2))\langle\hat{\eta}_c\rangle - 2\varepsilon\kappa(\kappa^2 - 2\varepsilon^2)\big\langle\hat{\sigma}_c + \hat{\sigma}^{\dagger}_c\big\rangle\bigg).
\end{eqnarray}
\indent
The three-level atom inside the closed cavity doesn't interact with the vacuum reservoir outside the cavity. Therefore, the equation of evolution of the density operator for this atom has the form [22]
\begin{equation}\label{75}
  \frac{d}{dt}\hat{\rho}(t) = -i\big[\hat{H}_a, \hat{\rho}\big].
\end{equation}
The equation of evolution for the expectation value of an atomic operator $\hat{\sigma}(t)$ can be written as
\begin{equation}\label{76}
  \frac{d}{dt}\langle \hat{\sigma}(t)\rangle = Tr\bigg(\frac{d}{dt}\hat{\rho}(t)\hat{\sigma}\bigg),
\end{equation}
so that in view of Eq. \eqref{75}, there follows
\begin{equation}\label{77}
  \frac{d}{dt}\langle \hat{\sigma}(t)\rangle = -iTr\bigg(\big[\hat{H}_a, \hat{\rho}\big]\hat{\sigma}\bigg).
\end{equation}
Now one can establish that
\begin{equation}\label{78}
  \frac{d}{dt}\langle \hat{\sigma}(t)\rangle = -i\langle\big[\hat{\sigma}, \hat{H}_a\big]\rangle.
\end{equation}
Applying this relation along with Eq. \eqref{7}, we readily get
\begin{equation}\label{79}
 \frac{d}{dt}\langle \hat{\sigma}_a(t)\rangle = g\bigg\langle \big(\hat{\eta}_b(t) - \hat{\eta}_a(t)\big)\hat{a}(t) + \hat{b}^{\dagger}(t)\hat{\sigma}_c(t)\bigg\rangle,
\end{equation}
\begin{equation}\label{80}
 \frac{d}{dt}\langle \hat{\sigma}_b(t)\rangle = g\bigg\langle -\hat{a}^{\dagger}(t)\hat{\sigma}_c(t) + \big(\hat{\eta}_c(t) - \hat{\eta}_b(t)\big)\hat{b}(t)\bigg\rangle,
\end{equation}
\begin{equation}\label{81}
 \frac{d}{dt}\langle \hat{\sigma}_c(t)\rangle = g\bigg\langle \hat{\sigma}_b(t)\hat{a}(t) - \hat{\sigma}_a(t)\hat{b}(t)\bigg\rangle,
\end{equation}
\begin{equation}\label{82}
 \frac{d}{dt}\langle \hat{\eta}_a(t)\rangle = g\bigg\langle \hat{\sigma}^{\dagger}_a(t)\hat{a}(t) + \hat{a}^{\dagger}(t)\hat{\sigma}_a(t)\bigg\rangle,
\end{equation}
\begin{equation}\label{83}
 \frac{d}{dt}\langle \hat{\eta}_b(t)\rangle = g\bigg\langle \hat{\sigma}^{\dagger}_b(t)\hat{b}(t) + \hat{b}^{\dagger}(t)\hat{\sigma}_b(t) - \big(\hat{\sigma}^{\dagger}_a(t)\hat{a}(t) + \hat{a}^{\dagger}(t)\hat{\sigma}_a(t)\big)\bigg\rangle,
\end{equation}
\begin{equation}\label{84}
 \frac{d}{dt}\langle \hat{\eta}_c(t)\rangle = -g\bigg\langle \hat{\sigma}^{\dagger}_b(t)\hat{b}(t) + \hat{b}^{\dagger}(t)\hat{\sigma}_b(t)\bigg\rangle,
\end{equation}
In views of Eqs. \eqref{27}, \eqref{28}, and \eqref{43} together with their adjoint, Eqs. \eqref{79}-\eqref{84} take the form
\begin{equation}\label{85}
 \frac{d}{dt}\langle\hat{\sigma}_a(t)\rangle = -\frac{\kappa^2\varepsilon\gamma_c}{\kappa^2 - 4\varepsilon^2}\bigg\langle \frac{\hat{\sigma}_a(t)}{2\varepsilon} - \frac{\hat{\sigma}^{\dagger}_b(t)}{\kappa}\bigg\rangle,
\end{equation}
\begin{equation}\label{86}
 \frac{d}{dt}\langle \hat{\sigma}_b(t)\rangle = -\frac{\kappa^2\gamma_c}{\kappa^2 - 4\varepsilon^2}\langle\hat{\sigma}_b(t)\rangle, 
\end{equation}
\begin{equation}\label{87}
 \frac{d}{dt}\langle \hat{\sigma}_c(t)\rangle = \frac{\kappa^2\varepsilon\gamma_c}{4\varepsilon^2 - \kappa^2}\bigg\langle\frac{\hat{\sigma}_c(t)}{2\varepsilon} + \frac{1}{\kappa}\big(\hat{\eta}_b(t) - \hat{\eta}_c(t)\big)\bigg\rangle,
\end{equation}
\begin{equation}\label{88}
 \frac{d}{dt}\langle \hat{\eta}_a(t)\rangle = \frac{\kappa^2\varepsilon\gamma_c}{4\varepsilon^2 - \kappa^2}\bigg\langle-\frac{1}{\kappa}\big(\hat{\sigma}^{\dagger}_c(t) + \hat{\sigma}_c(t)\big) + \frac{\hat{\eta}_a}{\varepsilon}\bigg\rangle,
\end{equation}
\begin{equation}\label{89}
 \frac{d}{dt}\langle \hat{\eta}_b(t)\rangle = \frac{\kappa^2\varepsilon\gamma_c}{4\varepsilon^2 - \kappa^2}\bigg\langle\frac{1}{\kappa}\big(\hat{\sigma}^{\dagger}_c(t) + \hat{\sigma}_c(t)\big) + \frac{1}{\varepsilon}\big(\hat{\eta}_b(t) - \hat{\eta}_a(t)\big)\bigg\rangle,
\end{equation}
\begin{equation}\label{90}
 \frac{d}{dt}\langle \hat{\eta}_c(t)\rangle = -\frac{\kappa^2\varepsilon\gamma_c}{4\varepsilon^2 - \kappa^2}\bigg\langle \frac{\hat{\eta}_b(t)}{\epsilon}\bigg\rangle,
\end{equation}
where $\gamma_c = \frac{4g^2}{\kappa} $ is the stimulated emission decay constant.\\
\indent
We find the steady-state solutions of Eqs. \eqref{87}-\eqref{89} to be of the form
\begin{equation}\label{91}
 \langle\hat{\sigma}_c\rangle = \frac{2\varepsilon}{\kappa}\bigg(\langle\hat{\eta}_c \rangle - \langle\hat{\eta}_b\rangle\bigg),
\end{equation}
\begin{equation}\label{92}
 \langle\hat{\sigma}^{\dagger}_c\rangle + \langle\hat{\sigma}_c\rangle = \frac{\kappa}{\varepsilon}\langle\hat{\eta}_a\rangle,
\end{equation}
\begin{equation}\label{93}
\langle\hat{\eta}_a\rangle - \langle\hat{\eta}_b\rangle = \frac{\varepsilon}{\kappa} \bigg(\langle\hat{\sigma}^{\dagger}_c\rangle + \langle\hat{\sigma}_c\rangle\bigg).
\end{equation}
Now from Eqs. \eqref{92} and \eqref{93}, one can easily get
\begin{equation}\label{94}
 \langle\hat{\eta}_b\rangle = 0.
\end{equation}
Upon substituting Eq. \eqref{91} into Eq. \eqref{92} and taking into account Eq.\eqref{94}, we readily obtain
\begin{equation}\label{95}
 \langle\hat{\eta}_c\rangle = \frac{\kappa^2}{4\varepsilon^2}\langle\hat{\eta}_a\rangle,
\end{equation}
The completeness relation for the three-level atom is given by
\begin{equation}\label{96}
\langle\hat{\eta}_a\rangle +  \langle\hat{\eta}_b\rangle +  \langle\hat{\eta}_c\rangle = 1,
\end{equation}
where $\langle\hat{\eta}_a\rangle$, $\langle\hat{\eta}_b\rangle$, and $\langle\hat{\eta}_c\rangle$ are the probabilities to find the atom in the top, intermediate, and bottom levels, respectively. Therefore, applying Eqs. \eqref{94} and \eqref{95},  we arrive at
\begin{equation}\label{97}
 \langle\hat{\eta}_a\rangle = \frac{4\varepsilon^2}{\kappa^2 + 4\varepsilon^2}
\end{equation}
and
\begin{equation}\label{98}
 \langle\hat{\eta}_c\rangle = \frac{\kappa^2}{\kappa^2 + 4\varepsilon^2}.
\end{equation}
Finally, combination of Eqs. \eqref{91}, \eqref{94}, and  \eqref{98} leads to
\begin{equation}\label{99}
 \langle\hat{\sigma}_c\rangle = \frac{2\varepsilon\kappa }{\kappa^2 + 4\varepsilon^2}.
\end{equation}
\section{Mean Photon Number}
\noindent
Next we calculate the mean photon number for the two-mode cavity light defined by
\begin{equation}\label{100}
 \hat{c} = \hat{a} + \hat{b}.
\end{equation}
The mean photon number for the two-mode cavity light is given by
\begin{equation}\label{101}
 \bar{n} = \langle\hat{c}^{\dagger}\hat{c}\rangle.
\end{equation}
Then on the basis of Eq. \eqref{100} and its adjoint, we arrive at
\begin{equation}\label{102}
 \bar{n} = \langle\hat{a}^{\dagger}\hat{a}\rangle + \langle\hat{b}^{\dagger}\hat{b}\rangle + \langle\hat{a}^{\dagger}\hat{b}\rangle + \langle\hat{b}^{\dagger}\hat{a}\rangle.
\end{equation}
On account of Eq. \eqref{67} and its complex conjugate, Eq. \eqref{102} takes the form
\begin{equation}\label{103}
 \bar{n} = \langle\hat{a}^{\dagger}\hat{a}\rangle + \langle\hat{b}^{\dagger}\hat{b}\rangle.
\end{equation}
We see that the mean photon number of the two-mode cavity light is the sum of the mean photon numbers of light modes a and b. Then taking into account Eqs. \eqref{71} and \eqref{72}, we get 
\begin{eqnarray}\label{104}
 \bar{n} &=& \ \frac{4\varepsilon^2}{\kappa^2 - 4\varepsilon^2} + \frac{\kappa\gamma_c}{(\kappa^2 - 4\varepsilon^2)^2}\bigg((2\varepsilon^2 + \kappa^2)\langle\hat{\eta}_a\rangle + (4\varepsilon^2 + \kappa^2)\langle\hat{\eta}_b\rangle + 2\varepsilon^2\langle\hat{\eta}_c\rangle \nonumber\\&& - 2\varepsilon\kappa\big\langle\hat{\sigma}_c + \hat{\sigma}^{\dagger}_c\big\rangle\bigg)
\end{eqnarray}
and employing Eqs. \eqref{94}, \eqref{97}, \eqref{98} and \eqref{99}, one readily obtains
\begin{eqnarray}\label{105}
 \bar{n} &=& \ \frac{2\varepsilon^2}{\kappa^2 - 4\varepsilon^2}\bigg[2 - \frac{\kappa\gamma_c}{\kappa^2 + 4\varepsilon^2}\bigg].
\end{eqnarray}
\begin{figure*}
 \centering
\begin{center}
\includegraphics[width=12cm, height = 10cm]{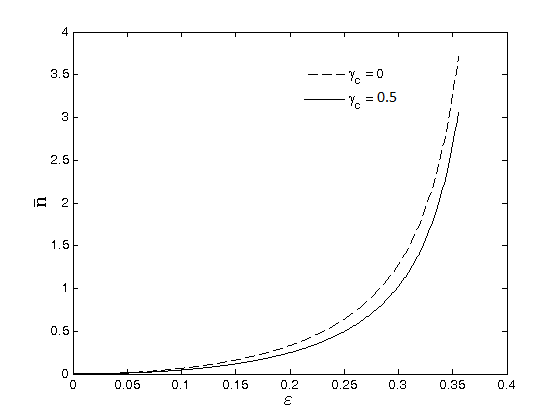}
\caption{Plots of the mean photon number [Eq. \eqref{105}] versus $\varepsilon$ for $\kappa = 0.8$.}
\end{center}
\end{figure*}
\indent
  We now consider the special case in which the subharmonic light modes don't interact with the three-level atom. Hence upon setting $\gamma_c = 0$, Eq. \eqref{105} reduces to [10]
\begin{equation}\label{106}
 \bar{n} = \frac{4\varepsilon^2}{\kappa^2 - 4\varepsilon^2}.
\end{equation}
This represents the mean photon number of the superposed subharmonic light modes in the absence of interaction with the three-level atom. From the plots in Fig.2 we observe that the effect of the interaction of the subharmonic light modes a and b with the three-level atom is to decrease the mean photon number of the two-mode cavity light.\\ 
\section{Quadrature Squeezing}
We next proceed to calculate the quadrature squeezing. To this end, applying the large-time approximation scheme to Eq. \eqref{86}, we get
\begin{equation}\label{107}
 \langle\hat{\sigma}_b(t)\rangle = 0.
\end{equation}
Then employing the complex conjugate of this result, one can put Eq. \eqref{85} in the form
\begin{equation}\label{108}
\frac{d}{dt} \langle\hat{\sigma}_a\rangle = -\frac{\kappa^2\gamma_c}{\kappa^2 - 4\varepsilon^2}\langle\hat{\sigma}_a(t)\rangle.
\end{equation}
With the atom considered to be initially in the bottom level, the solution of Eq. \eqref{108} turns out to be
\begin{equation}\label{109}
\langle\hat{\sigma}_a(t)\rangle = 0.
\end{equation}
Moreover, with the atom considered to be initially in the bottom level, the solution of Eq. \eqref{86} is found to be
\begin{equation}\label{110}
\langle\hat{\sigma}_b(t)\rangle = 0.
\end{equation}
Now substituting the complex conjugate of the expectation value of Eq. \eqref{28} into Eq. \eqref{11}, we get
\begin{equation}\label{111}
 \frac{d}{dt}\langle\hat{a}(t)\rangle = -\bigg(\frac{\kappa^2 - 4\varepsilon^2}{2\kappa}\bigg)\langle\hat{a}(t)\rangle + g \bigg( \frac{2\varepsilon}{\kappa}\langle\hat{\sigma}^{\dagger}_b(t)\rangle - \langle\hat{\sigma}_a(t)\rangle\bigg)
\end{equation}
and on taking into account Eq. \eqref{107}, we have
\begin{equation}\label{112}
 \frac{d}{dt}\langle\hat{a}(t)\rangle = -\bigg(\frac{\kappa^2 - 4\varepsilon^2}{2\kappa}\bigg)\langle\hat{a}(t)\rangle - g\langle\hat{\sigma}_a(t)\rangle.
\end{equation}
\indent
The solution of this equation is a well-behaved function provided that 
\begin{equation}\label{113}
 \frac{\kappa^2 - 4\varepsilon^2}{2\kappa} > 0.
\end{equation}
It then follows that
\begin{equation}\label{114}
 \varepsilon < \frac{\kappa}{2}.
 \end{equation}
We realize that the the solution of Eq. \eqref{112} is expressible for $\varepsilon < \frac{\kappa}{2}$ as
\begin{eqnarray}\label{115}
 \langle\hat{a}(t)\rangle &=& \langle\hat{a}(0)\rangle exp\bigg(-\big(\frac{\kappa^2 - 4\varepsilon^2}{2\kappa}\big)t\bigg)\nonumber\\&& - g\int^{t}_{0} exp\bigg(-\big(\frac{\kappa^2 - 4\varepsilon^2}{2\kappa}\big)(t-t')\bigg)\langle\hat{\sigma}_a(t')\rangle dt'.
\end{eqnarray}
Hence with the assumption that the cavity light is initially in a vacuum state and in view of Eq. \eqref{109}, Eq. \eqref{115} turns out to be 
\begin{equation}\label{116}
\langle\hat{a}(t)\rangle = 0.
\end{equation}
Applying the large-time approximation scheme to Eq. \eqref{85}, we obtain
\begin{equation}\label{117}
\langle\hat{\sigma}_a(t)\rangle = \frac{2\varepsilon}{\kappa}\langle\hat{\sigma}^{\dagger}_b(t)\rangle.
\end{equation}
Using the complex conjugate of the expectation value of Eq. \eqref{27} along with Eq. \eqref{117}, one can write Eq.\eqref{12} as
\begin{equation}\label{118}
\frac{d}{dt}\langle\hat{b}(t)\rangle = -\bigg(\frac{\kappa^2 - 4\varepsilon^2}{2\kappa}\bigg)\langle\hat{b}(t)\rangle- \frac{g}{\kappa^2}\bigg(\kappa^2 - 2\varepsilon^2\bigg)\langle\hat{\sigma}_b(t)\rangle.
\end{equation}
In view of Eq. \eqref{110} and the assumption that the cavity light is initially in a vacuum state, the solution of Eq. \eqref{118} turns out to be
\begin{equation}\label{119}
\langle\hat{b}(t)\rangle = 0.
\end{equation}
On account of Eqs. \eqref{116} and \eqref{119}, Eq. \eqref{100} takes the form
\begin{equation}\label{120}
\langle\hat{c}(t)\rangle = 0.
\end{equation}
\indent
The squeezing properties of the two-mode cavity light are described by two quadrature operators defined by
\begin{equation}\label{121}
 \hat{c}_+ = \hat{c}^{\dagger} + \hat{c}
\end{equation}
and
\begin{equation}\label{122}
 \hat{c}_- = i(\hat{c}^{\dagger} - \hat{c}).
\end{equation}
Applying Eqs. \eqref{121} and \eqref{122}, it can be readily established that
\begin{equation}\label{123}
 \big[\hat{c}_+, \hat{c}_-\big] = 2i\big[\hat{c}, \hat{c}^{\dagger}\big].
\end{equation}
It then follows that [22]
\begin{equation}\label{124}
 \Delta c_-\Delta c_+ \geq \big|\big\langle\big[\hat{c}, \hat{c}^{\dagger}\big]\big\rangle \big|.
\end{equation} 
Now we proceed to obtain the explicit form of the commutator $\big[\hat{c}, \hat{c}^{\dagger}\big]$. To this end, we note that
\begin{equation}\label{125}
 \hat{c}\hat{c}^{\dagger} = \hat{a}\hat{a}^{\dagger} + \hat{b}\hat{b}^{\dagger}
\end{equation}
and in view of Eqs. \eqref{73} and \eqref{74}, we get 
\begin{eqnarray}\label{126}
\hat{c}\hat{c}^{\dagger}  &=& 2 + \frac{4\varepsilon^2}{\kappa^2 - 4\varepsilon^2} + \frac{\kappa\gamma_c}{(\kappa^2 - 4\varepsilon^2)^2}\bigg(6\varepsilon^2\hat{\eta}_a + (4\varepsilon^2 + \kappa^2)\hat{\eta}_b + (\kappa^2 -2\varepsilon^2)\hat{\eta}_c\nonumber\\&& - 2\varepsilon\kappa\big(\hat{\sigma}_c + \hat{\sigma}_c\big)\bigg).
\end{eqnarray}
Employing Eqs. \eqref{104} and \eqref{126}, we find
\begin{equation}\label{127}
\big[\hat{c}, \hat{c}^{\dagger}\big] = 2 + \frac{\kappa\gamma_c}{\kappa^2 - 4\varepsilon^2}\bigg(\hat{\eta}_c - \hat{\eta}_a\bigg).
\end{equation} 
We identify the first and second terms in the above equation to be due to the ordering of the noise and atomic operators.
On account of Eq. \eqref{127}, Eq. \eqref{124} takes the form
\begin{equation}\label{128}
 \Delta c_-\Delta c_+ \geq \bigg|2 + \frac{\kappa\gamma_c}{\kappa^2 - 4\varepsilon^2}\bigg(\langle\hat{\eta}_c\rangle - \langle\hat{\eta}_a\rangle\bigg)\bigg|.
\end{equation}
Now substituting Eqs. \eqref{97} and\eqref{98} into Eq. \eqref{128}, one readily finds
\begin{equation}\label{129}
 \Delta c_-\Delta c_+ \geq \bigg|2 + \frac{\kappa\gamma_c}{\kappa^2 + 4\varepsilon^2}\bigg|.
\end{equation} 
\indent
The variance of the quadrature operators is expressible as
 \begin{equation}\label{130}
  (\Delta c\pm)^2 = \pm\langle(\hat{c}^{\dagger} \pm \hat{c})^2\rangle \mp(\langle\hat{c}^{\dagger}\rangle \pm \langle\hat{c}\rangle)^2
 \end{equation}
and on account of Eq. \eqref{120}, we have
\begin{equation}\label{131}
  (\Delta c\pm)^2 = \langle\hat{c}^{\dagger}\hat{c}\rangle + \langle\hat{c}\hat{c}^{\dagger}\rangle \pm\big( \langle\hat{c}^{\dagger2}\rangle + \langle\hat{c}^2\rangle\big).
 \end{equation}
Applying Eq. \eqref{100} along with Eq. \eqref{68}, one readily establishes that
\begin{equation}\label{132}
 \langle\hat{c}^2\rangle = \big\langle\hat{a}\hat{b} + \hat{b}\hat{a}\big\rangle,
 \end{equation} 
so that on the basis of Eqs. \eqref{58} and \eqref{59}, this expression turns out to be
\begin{eqnarray}\label{133}
\langle\hat{c}^2\rangle = -\frac{2\varepsilon}{\kappa}\big\langle\hat{a}^{\dagger}\hat{a} + \hat{b}^{\dagger}\hat{b}\big\rangle - \frac{4g^2}{\kappa^2 - 4\varepsilon^2}\bigg[\frac{2\varepsilon\langle\hat{\eta}_b\rangle}{\kappa} + \frac{\varepsilon\langle\hat{\eta}_a + \hat{\eta}_c\rangle}{\kappa}  -\langle\hat{\sigma}_c\rangle\bigg] - \frac{2\varepsilon}{\kappa}.
\end{eqnarray}
On account of Eqs. \eqref{71} and \eqref{72}, there follows
\begin{eqnarray}\label{134}
\langle\hat{c}^2\rangle &=& -\frac{2\varepsilon}{\kappa} - \frac{2\varepsilon}{\kappa}\bigg(\frac{4\varepsilon^2}{\kappa^2 - 4\varepsilon^2}\bigg) - \frac{\kappa\gamma_c}{(\kappa^2 - 4\varepsilon^2)^2}\bigg(3\varepsilon\kappa\langle\hat{\eta}_a\rangle + 4\varepsilon\kappa\langle\hat{\eta}_b\rangle + \varepsilon\kappa\langle\hat{\eta}_c\rangle \nonumber\\&&- [4\varepsilon^2\langle\hat{\sigma}_c + \hat{\sigma}^{\dagger}_c\rangle + ( \kappa^2 -4\varepsilon^2)\langle\hat{\sigma}_c\rangle]\bigg)
\end{eqnarray}
In a similar manner, one easily gets
\begin{eqnarray}\label{135}
\langle\hat{c}^{\dagger2}\rangle &=& -\frac{2\varepsilon}{\kappa} - \frac{2\varepsilon}{\kappa}\bigg(\frac{4\varepsilon^2}{\kappa^2 - 4\varepsilon^2}\bigg) - \frac{\kappa\gamma_c}{(\kappa^2 - 4\varepsilon^2)^2}\bigg(3\varepsilon\kappa\langle\hat{\eta}_a\rangle + 4\varepsilon\kappa\langle\hat{\eta}_b\rangle + \varepsilon\kappa\langle\hat{\eta}_c\rangle \nonumber\\&&- [4\varepsilon^2\langle\hat{\sigma}_c + \hat{\sigma}^{\dagger}_c\rangle + ( \kappa^2 -4\varepsilon^2)\langle\hat{\sigma}^{\dagger}_c\rangle]\bigg)
\end{eqnarray}
\begin{figure*}
 \centering
 \begin{center}
 \includegraphics[width=12cm, height = 10cm]{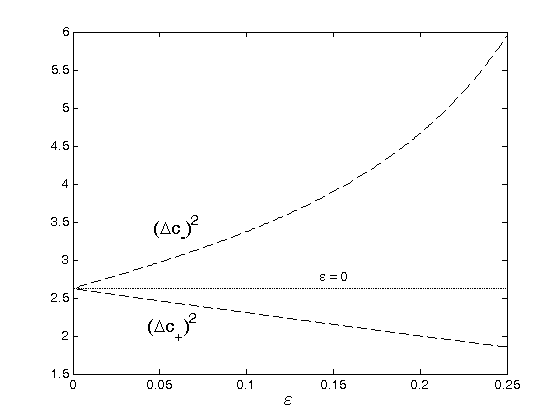}
  \caption{Plots of the plus and minus quadrature variances [Eqs. \eqref{138} and \eqref{139}] versus $\varepsilon$ for $\kappa = 0.8$ and $\gamma_c = 0.5$.}
 \end{center}
\end{figure*}\\
Upon substituting Eqs. \eqref{104}, \eqref{126}, \eqref{134}, and \eqref{135} into Eq. \eqref{131}, one obtains
\begin{eqnarray}\label{136}
(\Delta c_+)^2 &=& 2 - \frac{4\varepsilon}{\kappa + 2\varepsilon} + \frac{\kappa\gamma_c}{(\kappa^2 - 4\varepsilon^2)^2}\bigg(\langle\hat{\eta}_a\rangle(8\varepsilon^2 + \kappa^2 - 6\varepsilon\kappa) + 2(4\varepsilon^2 + \kappa^2 - 4\varepsilon\kappa)\langle\hat{\eta}_b\rangle\nonumber\\&& + (\kappa -2\varepsilon)\kappa \langle\hat{\eta}_c\rangle + (\kappa(\kappa- 4\varepsilon) + 4\varepsilon^2)\langle\hat{\sigma}_c + \hat{\sigma}^{\dagger}_c\rangle\bigg)
\end{eqnarray}
and
 \begin{eqnarray}\label{137}
(\Delta c_-)^2 &=& 2 + \frac{4\varepsilon}{\kappa - 2\varepsilon} + \frac{\kappa\gamma_c}{(\kappa^2 - 4\varepsilon^2)^2}\bigg(\langle\hat{\eta}_a\rangle(8\varepsilon^2 + \kappa^2 + 6\varepsilon\kappa) + 2(4\varepsilon^2 + \kappa^2 + 4\varepsilon\kappa)\langle\hat{\eta}_b\rangle\nonumber\\&& + (\kappa + 2\varepsilon)\kappa \langle\hat{\eta}_c\rangle -(\kappa(\kappa+ 4\varepsilon) + 4\varepsilon^2)\langle\hat{\sigma}_c + \hat{\sigma}^{\dagger}_c\rangle\bigg).
\end{eqnarray}
\noindent
Furthermore, on substituting Eqs. \eqref{94}, \eqref{97}, \eqref{98} and \eqref{99} into Eqs. \eqref{136} and \eqref{137}, one readily finds
\begin{eqnarray}\label{138}
(\Delta c_+)^2 &=& 2 - \frac{4\varepsilon}{\kappa + 2\varepsilon} + \frac{\kappa\gamma_c}{(\kappa^2 + 4\varepsilon^2)(\kappa^2 - 4\varepsilon^2)^2}\bigg[4\varepsilon^2\bigg(8\varepsilon^2 + \kappa^2 - 6\varepsilon\kappa\bigg) + \kappa^3(\kappa - 2\varepsilon)\nonumber\\&& + 4\varepsilon\kappa\bigg(\kappa(\kappa - 4\varepsilon)+ 4\varepsilon^2\bigg)\bigg]
\end{eqnarray}
and
 \begin{eqnarray}\label{139}
(\Delta c_-)^2 &=& 2 + \frac{4\varepsilon}{\kappa - 2\varepsilon} + \frac{\kappa\gamma_c}{(\kappa^2 + 4\varepsilon^2)(\kappa^2 - 4\varepsilon^2)^2}\bigg[4\varepsilon^2\bigg(8\varepsilon^2 + \kappa^2 + 6\varepsilon\kappa\bigg) + \kappa^3(\kappa + 2\varepsilon)\nonumber\\&& - 4\varepsilon\kappa\bigg(\kappa(\kappa + 4\varepsilon)+ 4\varepsilon^2\bigg)\bigg].
\end{eqnarray}
We note that for $\varepsilon = 0$, Eqs. \eqref{138} and \eqref{139} reduce to
 \begin{equation}\label{140}
  (\Delta c_+)_{v}^2 = (\Delta c_-)_{v}^2 = 2 + \frac{\gamma_c}{\kappa} .
 \end{equation}
This indeed represents the quadrature variance of a two-mode cavity vacuum state, with the effect of the noise operators included. From the plots in Fig.3, we observe that the two-mode cavity light is in a squeezed state and the squeezing occurs in the plus quadrature. 
\noindent

Now we calculate the quadrature squeezing of the two-mode cavity light relative to the quadrature variance of the two-mode cavity vacuum state. We define the quadrature squeezing of the two-mode cavity light by [10]
\begin{equation}\label{141}
 S = \frac{(\Delta c_+)^2_v - (\Delta c_+)^2}{(\Delta c_+)^2_v}.
\end{equation}
\begin{figure*}
\centering
\begin{center}
\includegraphics[width=12cm, height = 10cm]{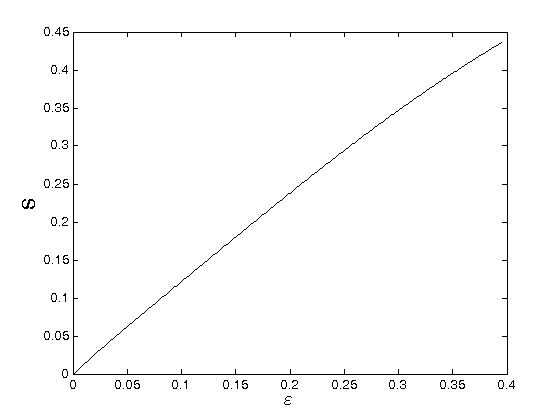}
\caption{Plots of quadrature squeezing [Eq. \eqref{142}] versus $\varepsilon$ for $\kappa = 0.8$ and $\gamma_c = 0.5$.}
\end{center}
\end{figure*}
Upon using Eqs. \eqref{138} and \eqref{140} in Eq. \eqref{141}, we find
 \begin{eqnarray}\label{142}
S &=& 1- \frac{\kappa}{\gamma_c + 2\kappa}\bigg[2 - \frac{4\varepsilon}{\kappa + 2\varepsilon} + \frac{\kappa\gamma_c}{(\kappa^2 + 4\varepsilon^2)(\kappa^2 - 4\varepsilon^2)^2}\bigg[4\varepsilon^2\bigg(8\varepsilon^2 + \kappa^2 - 6\varepsilon\kappa\bigg)\nonumber\\&& + \kappa^3(\kappa - 2\varepsilon) + 4\varepsilon\kappa\bigg(\kappa(\kappa - 4\varepsilon)+ 4\varepsilon^2\bigg)\bigg]\bigg]
 \end{eqnarray}
We now consider the special case in which the three-level atom does not interact with the light modes a and b. Hence upon setting $\gamma_c = 0$, Eq. \eqref{142} reduces to [10]
\begin{equation}\label{143}
 S = \frac{2\varepsilon}{\kappa + 2\varepsilon}.
\end{equation}
This represents the quadrature squeezing of the superposed subharmonic light modes. We realize from the plot in Fig.4 that the effect of the interaction of the subharmonic light modes with the three-level atom  is to decrease the quadrature squeezing of the two-mode cavity light.\\

\section{Conclusion}
\noindent 
We have studied the interaction of the subharmonic light modes, emerging from a nonlinear crystal driven by coherent light, with a three-level atom. Applying the steady-state solutions of the pertinent equations of evolution, we have calculated the mean photon number and the quadrature squeezing of the two-mode cavity light available following this interaction. We have found that the effect of this interaction is to decrease the quadrature squeezing and the mean photon number of the two-mode cavity light.

\end{document}